\documentclass[reprint,superscriptaddress,floatfix,prb]{revtex4-1}

\usepackage{amsmath}
\usepackage{amssymb}
\usepackage{float}

\usepackage{graphicx}
\usepackage{dcolumn}
\usepackage{bm}

\newcommand{\colvec}[2][.8]{%
  \scalebox{#1}{%
    \renewcommand{\arraystretch}{.8}%
    $\begin{pmatrix}#2\end{pmatrix}$%
  }
}

\begin{document}

\newcolumntype{L}[1]{>{\raggedright\arraybackslash}p{#1}}
\newcolumntype{C}[1]{>{\centering\arraybackslash}p{#1}}
\newcolumntype{R}[1]{>{\raggedleft\arraybackslash}p{#1}}

\title{Broadband Circular Polarization Time-Domain Terahertz Spectroscopy}
\author{Evan. V. Jasper}
\email{jasper.31@osu.edu}
\affiliation{Department of Physics, The Ohio State University. Columbus, OH 43210, USA}
\author{T. T. Mai}
\affiliation{Department of Physics, The Ohio State University. Columbus, OH 43210, USA}
\affiliation{Nanoscale Device Characterization Division, Physical Measurement Laboratory, NIST. Gaithersburg, MD 20899, USA}
\author{M. T. Warren}
\affiliation{Department of Physics, The Ohio State University. Columbus, OH 43210, USA}
\affiliation{Nationwide Insurance. Columbus, OH 43215, USA}
\author{R. K. Smith}
\affiliation{Department of Physics, The Ohio State University. Columbus, OH 43210, USA}
\author{D. M. Heligman}
\affiliation{Department of Physics, The Ohio State University. Columbus, OH 43210, USA}
\author{E. McCormick}
\affiliation{Department of Physics, The Ohio State University. Columbus, OH 43210, USA}
\affiliation{Metron, Incorporated. Reston, VA 20190, USA}
\author{Y.S. Ou}
\affiliation{Department of Physics, The Ohio State University. Columbus, OH 43210, USA}
\affiliation{University of Delaware, Department of Material Science and Engineering. Newark, DE 19716, USA}
\author{M. Sheffield}
\affiliation{Department of Physics, The Ohio State University. Columbus, OH 43210, USA}
\affiliation{Department of Physics and Astronomy, University of Utah. Salt Lake City, UT 84112, USA}
\author{R. Vald\'es Aguilar}
\email{valdesaguilar.1@osu.edu}
\affiliation{Department of Physics, The Ohio State University. Columbus, OH 43210, USA}
\date{\today}




\begin{abstract}
Light-matter interactions are key in providing fundamental information about materials. Even in the linear-response regime, the spectroscopic response of a material encodes in it many properties of the ground state as well as of its excitations. This knowledge has been critical in our understanding of novel quantum materials, and the further improvement and extensions of linear spectroscopy will continue to be key in the exploration of novel states of matter. We report the development of broadband circular polarization spectroscopy in the terahertz range of the electromagnetic spectrum. We take advantage of a recent design of a broadband quarter wave plate, based on the Fresnel rhomb concept, and use it in conjunction with a polarization modulation technique to provide direct information of the response of a material to circularly polarized THz radiation; a new capability shown here for the first time. As an example of this technique we study the cyclotron resonance of a 2D electron gas from a AlGaAs--GaAs quantum well. We demonstrate the unique advantages that this technique will bring in the study of novel quantum materials.
\end{abstract}
\maketitle

\section{Introduction}
The terahertz (THz) frequency range is a critically important region of the electromagnetic spectrum where the electronic properties of many quantum materials have resonant responses. One terahertz corresponds to an energy of 4 meV ($\sim$50 K), and terahertz spectroscopy covers the energy range from $\sim$0.5 meV to 40 meV (0.1 $-$ 10 THz) \cite{nuss1998terahertz,jepsen2011terahertz,kampfrath2013resonant}. The rates of inelastic processes in solids also lie in this same frequency range, such as tunneling and quasiparticle scattering \cite{RMP.83.471}. Additionally, antiferromagnetic spin wave gaps are common in this frequency range \cite{sievers1963far}. In spite of this importance, progress using THz spectroscopy has been hindered by the lack of suitable sources and detectors. 
Although first invented in the 1970s \cite{auston1975picosecond,auston1980picosecond}, only in the last ten years has the technique of time-domain terahertz spectroscopy (TDTS) become robust enough to start filling in this gap, achieving its full maturity and reliability. Now, TDTS is a broadband, stable, and sensitive probe of quantum materials.

\begin{figure}[t!]
\includegraphics[width=1\columnwidth]{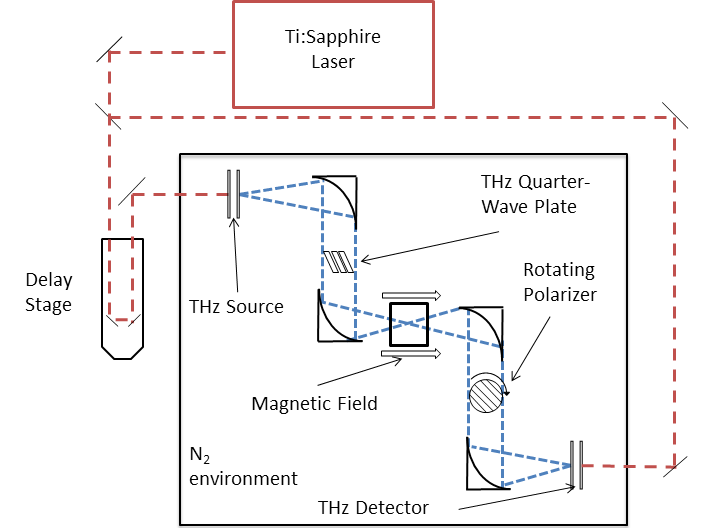}
\centering
\caption{Schematic of the time-domain terahertz spectroscopy setup. The Ti:Sapphire laser produces $\sim$ 100 fs long pulses of infrared radiation centered at 800 nm. The THz source and detector are a pair of photoconductive antennae grown on LT-GaAs. The entirety of the THz beam path is enclosed in an inert N$_2$ atmosphere to avoid water vapor absorption. The THz quarter wave plate (QWP) array described here, and the rotating polarizer device are also shown in a typical characterization configuration. The GaAs-AlGaAs quantum well was placed inside a magneto-optical cryostat in the Faraday geometry.}
\label{setup}
\end{figure}

The basic components of a transmission TDTS system are outlined in figure \ref{setup}. This system has been used in previous work \cite{ThucSFSO,MattGVS,pottscorrective}. Here an infrared laser (typically based on a Ti-doped sapphire crystal) generates pulses of radiation with temporal durations in the range of tens of femtoseconds. These pulses are split into two paths, one that goes to the emitter of the THz pulse, and the other goes to the detector of the THz pulse. Several options exist to generate and detect these THz pulses \cite{jepsen2011terahertz}. The two most common ones are based on photoconductive switches and nonlinear optical processes in inversion asymmetric crystals. Because both generator and detector are activated by the same infrared pulse, split into two paths, this technique is therefore coherent and sensitive to the electric field of the THz wave.
As the THz pulse travels through the system and the sample, it takes a certain amount of time to reach the detector; at the same time the infrared pulse that activates the detector has a variable time-delay line that adjusts the length of this path to match the travel time of the THz pulse. Thus, as the time delay is controlled, the detector maps out the waveform of the electric field of the THz pulse as a function of time. Because this technique measures the electric field, and not its intensity, it is sensitive to changes in phase introduced by the sample. It is this ability to detect the full complex optical response of a material that makes TDTS a particularly powerful tool in the study of materials.

Although polarization sensitive techniques have been developed for use with single frequency sources, such as far-infrared lasers \cite{vcerne2003measurement,SATIRACHAT2011}, these techniques are incompatible with the broadband nature of TDTS. We do however build on previous work in monochromatic polarization sensitive detection. In that technique, a QWP designed for the laser's single wavelength is made to spin at a high angular frequency, and as the laser beam passes through the spinning QWP, the polarization state of the light changes from linear to perfectly circularly polarized of both handedness in a continuous manner. 
By using phase sensitive lock-in detection, the modulated signal contains the information of the transmission coefficient for both right and left-handed circular polarization states of the incoming laser beam. This provided a highly sensitive way of detecting small changes of the polarization state induced by a sample, such as small Faraday rotation angles, or circular birefringence. Even with the high sensitivity of this method, its limitation lies in the fact that only a single frequency is measured at a time. Thus the spectral response cannot be readily obtained, unless different lasers and wave plates are used for each wavelength.

We detail here the construction of a pair of Fresnel rhomb arrays which can be inserted into a collimated section of nearly any THz beam line. We then proceed to combine those with a rotating polarizer \cite{morris2012polarization}. We also show that, with properly ordered and configured optical elements, we can directly measure all four components of the complex transmission matrix in the circular polarization basis using only two measurements. Finally, we demonstrate this capability with a system known to react directly to circularly polarized light, a two-dimensional electron gas (2DEG) in a AlGaAs-GaAs quantum well.

\section{Circular Polarization \& Jones matrices}

The development of a polarization modulation method combined with TDTS came to fruition only recently \cite{george2012terahertz,morris2012polarization}. This is based on a rapidly spinning element, in this case a linear THz polarizer, and is able to simultaneously detect the complex response of two elements in the transmission matrix. Here the transmission matrix (or $T$-matrix) is $T = \left(\begin{smallmatrix}a & b\\ c & d\end{smallmatrix}\right)$. The elements $a$, $b$, $c$, and $d$ are, in general, frequency-dependent complex values representing the transmission coefficients of the sample $E_{out}=T\cdot E_{in}$ . The calculation of the optical response for a particular experimental configuration can be done using the formalism of Jones calculus \cite{armitage2014constraints}. In this case the state of the polarization of light is represented with a two-row vector $E=\left(\begin{smallmatrix}E_x \\ E_y\end{smallmatrix}\right)$, and optical elements with two by two matrices; several examples of the matrices of these components are shown in table \ref{table1}. This technique of polarization modulation when combined with lock-in detection has been used to study topological insulators \cite{RolandoKerr,wu2016quantized}, the quantum Hall state in GaAs \cite{stier2015terahertz}, and also High-Tc superconductors \cite{lubashevsky2014optical}.

\begin{table*}
\centering
\begin{tabular}{| c | c| }
\hline
T-matrix in \break linear basis: & T-matrix in \break circular basis:\\[1pt]
$T=\colvec[.9]{a & b \\ c &  d}$ &  $T=\frac{1}{2}\colvec[.9]{a+d+i(b-c) & a-d-i(b+c) \\ a-d+i(b+c) & a+d-i(b-c)}$\\ [5pt]
\hline
Linear Polarizer along x-axis:  & QWP fast axis \break along x-axis:\\[1pt]
$P_x=\colvec[.9]{1 & 0 \\ 0 &  0 }$ & $QWP_x=e^{i\pi/4}\colvec[.9]{1 & 0 \\ 0 & i }$ \\[5pt]
\hline
Linear Polarizer along y-axis: & QWP fast axis \break along y-axis:\\[1pt]
$P_y=\colvec[.9]{ 0 & 0 \\ 0 & 1 }$ & $QWP_y=e^{i\pi/4}\colvec[.9]{1 & 0 \\ 0 & -i }$ \\[5pt]
\hline
Generic Linear Polarizer: & Generic QWP: \\[1pt]
$P_\theta=\colvec[.9]{\cos^2\theta & \sin\theta\cos\theta \\ \sin\theta\cos\theta & \sin^2\theta}$ & $QWP_\theta=\frac{1}{\sqrt{2}}\colvec[.9]{i+\cos2\theta & 2\cos\theta\sin\theta \\ 2\cos\theta\sin\theta & i-\cos2\theta}$\\[5pt]
\hline
Linearly polarized field: & Right circularly polarized field: \\[1pt]
$E_x=\colvec[.9]{1 \\ 0}$ & $E_+=\frac{1}{\sqrt{2}}\colvec[.9]{1 \\ -i}$\\[5pt]
\hline
Linearly polarized field: & Left circularly polarized field:\\[1pt]
$E_y=\colvec[.9]{0 \\ 1}$ & $E_-=\frac{1}{\sqrt{2}}\colvec[.9]{1 \\ i}$\\[5pt]
\hline
\end{tabular}
\caption{\label{tab:jones}The Jones matrices corresponding to various optical elements. With the exception of the $T$-matrix in the circular basis, all elements here are written in the linear basis with basis vectors $E_x$ and $E_y$. In the circular basis, the $T$-matrix is written in terms of the linear basis T-matrix elements.}\
\label{table1}
\end{table*}

The technique works in a configuration represented by the equation
\begin{eqnarray}
E_{out}=P_x\cdot P_{\Omega t}\cdot T\cdot E_x,
\end{eqnarray}
which means that a linearly polarized wave along the x-axis comes into the sample $T$, after transmission through the sample it passes to the rotating polarizer with a mechanical angular frequency $\Omega$, and then the beam goes through a linear polarizer along the x-axis. This results in 
\begin{eqnarray}
E=\frac{1}{2}\left(\begin{matrix}a+a\cos(2\,\Omega\, t)+c\sin(2\,\Omega \,t) \\ 0\end{matrix}\right)
\label{eq:E_rp_out}.
\end{eqnarray}
The modulation at twice the mechanical angular frequency is due to the 2-fold rotational symmetry of the linear polarizer, i.e. two positions with the same polarization orientation, as can be seen from the expression of $P_\theta$ in table \ref{tab:jones}. 
When this signal is fed to a phase sensitive lock-in amplifier that receives the modulation frequency 2$\Omega$ separately, it multiplies this reference with the signal $E_{out}$ and integrates the product over a predefined amount of time (the time constant). This will result in a two component signal made up of in-phase ($X$) and out-of-phase ($Y$) parts with respect to the reference. These two components are then recorded separately and are given by $X = a/2$, and $Y=c/2$. Thus, in a single experiment both the $a$ and $c$ components of the $T$-matrix are measured concurrently. If the configuration is modified to now detect the $y$ components, 
\begin{eqnarray}
E_{out}=P_y\cdot P_{\Omega t}\cdot T\cdot E_y
\label{eq:E_out},
\end{eqnarray}
then we would obtain $X = -d/2$, and $Y = b/2$. Therefore, two measurements are all that are needed to fully characterize the $T$-matrix of a sample using this polarization modulation technique \cite{morris2012polarization}.

Even though the power of this technique is readily apparent in the measurement of polarization-dependent phenomena in condensed matter, it lacks direct sensitivity to the circular polarization state of an incoming THz wave. There are examples, however, when it is possible to extract the circular polarization response from the linear ones \cite{LauritaHoMnO3,cheng2019magnetoterahertz,konocavityqed}. In some circumstances the sample may have a more physically relevant response to electromagnetic radiation that is circularly polarized. This is the case, for example, when the state of the sample breaks time-reversal symmetry, its more natural optical response will be in the difference in the complex indices of refraction for right and left circular polarization. For instance, a $T$-matrix of the form
\begin{eqnarray}
T=\left(\begin{matrix}a & b\\ -b & a\end{matrix}\right)
\label{eq:T_c4mag}
\end{eqnarray}
coincides with, among other things, the linear-basis representation of a four-fold symmetric material in a magnetic field parallel to the axis of the light propagation. In the circular basis, this $T$-matrix becomes diagonal with
\begin{eqnarray}
T=\left(\begin{matrix}a+ib & 0\\ 0 & a-ib\end{matrix}\right)=\left(\begin{matrix}t_{++} & 0\\ 0 & t_{--}\end{matrix}\right)
\label{eq:T_c4mag_cp}.
\end{eqnarray}
It is in this circular basis where the normal modes of propagation correspond to circularly polarized waves, so having available a broadband circularly polarized THz source would be advantageous in the study of quantum materials that break time reversal symmetry, as well as in the presence of an external magnetic field.

\section{Fresnel rhomb}
\begin{figure*}[t!b!h!]
\includegraphics[width=2\columnwidth]{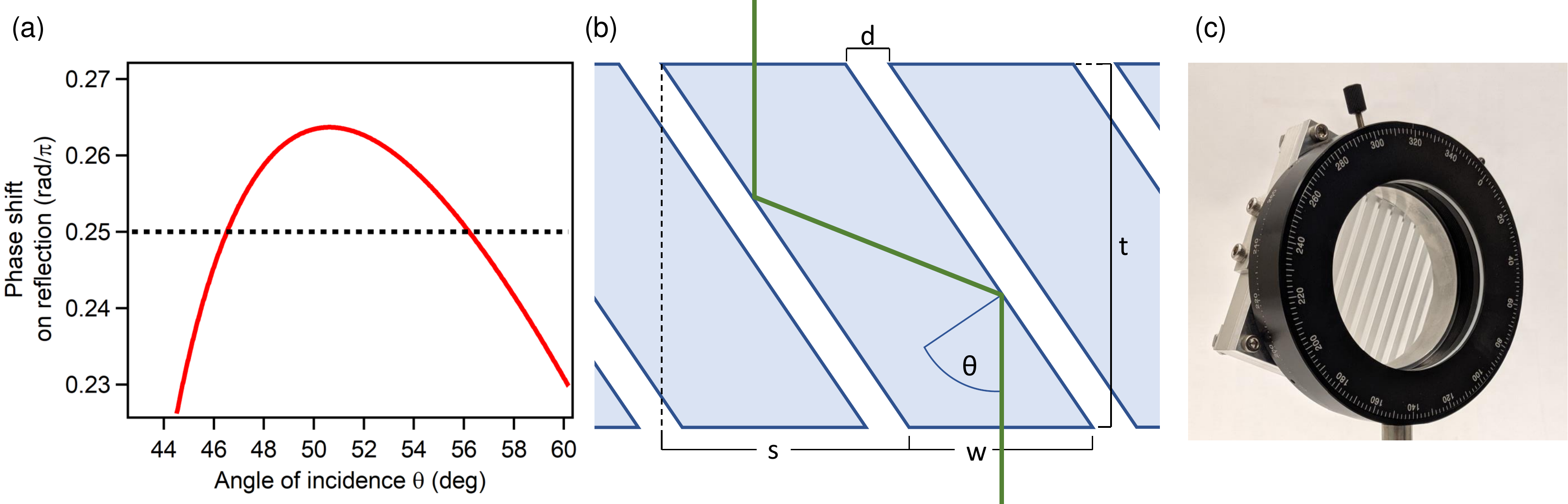}
\centering
\caption{(a) Phase shift of the THz radiation as a function of angle of incidence. $\pi/4$ radians is marked by the dotted line. There are two intersections between these lines showing the two possibilities of rhomb angle. (b) Diagram of the Fresnel rhomb array. Design parameters for our rhombs: rhomb separation d = 1 mm, rhomb thickness t = 11.7 mm, rhomb width w = 5.8 mm, lateral beam shift s = 8.7 mm, and angle of incidence $\theta$ = 55.67$^\circ$. (c) Photograph of the Fresnel rhomb array.}
\label{fig:rhomb}
\end{figure*}

There have been a series of demonstrations of broadband generation of circularly polarized THz pulses \cite{shan2009circularly,sato2013terahertz,kawada2014achromatic}, however some required an expensive and highly specialized piece of equipment to function \cite{sato2013terahertz}, and others use materials with high absorption loss for the THz pulse which reduces the efficiency and sensitivity of the technique \cite{shan2009circularly}. Our circular polarizer is based on an array of Fresnel rhombs---a prism device invented in the early 1800s that functions by delaying the phase of one of the linear components of light upon total internal reflection inside the prism \cite{fowles1989introduction}. Light enters the rhomb at an angle such that equal components of $s$-- and $p$--polarized THz impinge upon the internal surface of the rhomb (at the interface between the rhomb and vacuum) as shown in \ref{fig:rhomb}(b). The angle of incidence is chosen to create the condition of total internal reflection. The phase difference between the $s$-- and $p$--polarized components is given by
\begin{eqnarray}
\Delta\varphi=2\arctan\left(\frac{\cos\theta\sqrt{\sin^2\theta-n^2}}{\sin^2\theta}\right)
\label{eq:phase_shift}.
\end{eqnarray}
Here, $\theta$ is the angle of incidence and $n=1/n_{rhomb}$, where $n_{rhomb}$ is the real part of the refractive index of the prism material. 

Since circular polarized light is defined by the $\pm\pi/2$ phase difference between the $s$-- and $p$--components, it is crucial that the rhombs be made out of a material that exhibits a constant index of refraction over the THz operating range. Cyclic olefin copolymer TOPAS\textsuperscript{\textregistered} has the properties of a very flat index of refraction of $\sim$1.52 over a frequency range from at least 0.1 THz to 10 THz \cite{cunningham2011broadband}. Using this index of refraction we find that an angle of either 47.28$^\circ$ or 55.67$^\circ$ will give the desired phase shift between $s$-- and $p$--polarization of $\pi/4$ per bounce within the rhomb. We chose the latter value to reduce erroneous phase shift resulting from imperfect machining of the rhombs, because the slope of the phase shift as a function of reflection angle is lower compared to 47.28$^\circ$ as shown in figure~\ref{fig:rhomb}(a). 

\begin{figure}[t!b!h!]
\includegraphics[width=.95\columnwidth]{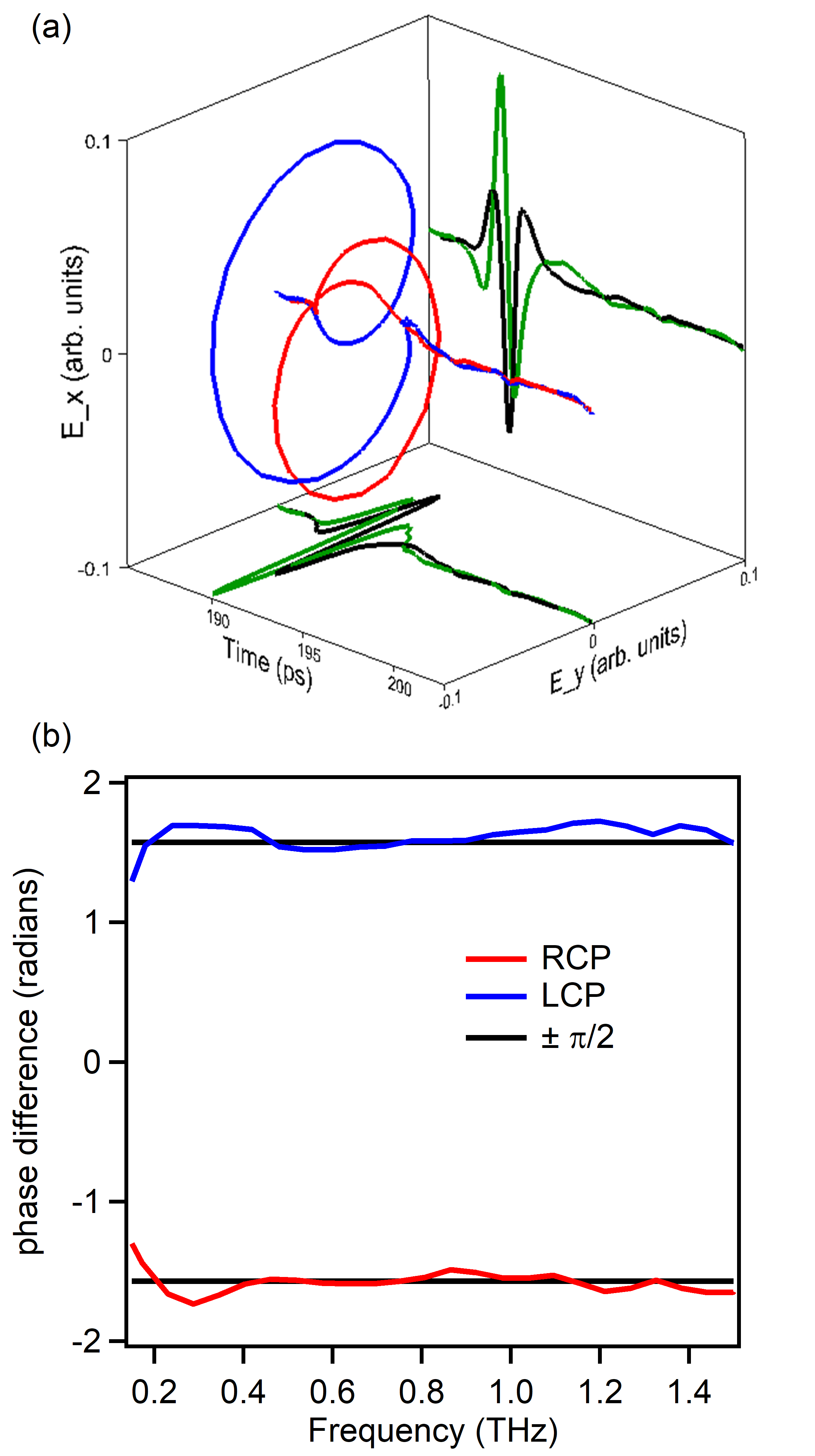}
\centering
\caption{(a) Time domain electric fields for left and right circularly polarized pulses. (b) Phase difference between horizontal and vertical components of the circularly polarized pulses for both LCP and RCP.}
\label{rhombpulse}
\end{figure}

The dimensions of the rhomb are key parameters in the rhomb's performance characteristics, and with the constraint of making an optical element capable of being inserted into any THz spectrometer, we have prioritized minimizing the path length through the Fresnel rhomb array. This minimized path length also reduces the lateral beam shift due to the array. Even though we made efforts to minimize path length through the array, absorption must still be considered. The absorption in TOPAS\textsuperscript{\textregistered} in our frequency range is very low, $\sim$1 cm$^{-1}$ \cite{cunningham2011broadband}, but not so low that we can ignore it. Other Fresnel rhomb array designs have been proposed and constructed prior to this work which have zero lateral beam shift at the cost of a much greater rhomb material thickness \cite{kawada2014achromatic}. Despite our efforts to reduce the path length in our rhomb, the dimensions of our rhomb lead to a path length in the rhomb of $\sim$1.9 cm. Though the path length could be reduced further by making each rhomb smaller, it would come at the expense of an increased number of rhombs in the array and, because of the need for finite gaps between each rhomb, a reduced cross section of the captured beam. See figure~\ref{fig:rhomb}(b). A photograph of one of the finished rhomb arrays is shown in figure~\ref{fig:rhomb}(c). It is mounted to allow for rotation around the optic axis.

The performance characteristics of our prism array are shown in figure \ref{rhombpulse}; these data were obtained using both the Fresnel rhomb array as a sample and the rotating polarizer setup in the following configuration 
\begin{eqnarray}
E_{out}=P_y\cdot P_{\Omega t}\cdot QWP_{\pm \pi/4}\cdot E_y
\label{eq:Eout_qwp}. 
\end{eqnarray}
Both circular polarizations can be generated with a single array by rotating the array by 90$^\circ$ around the axis of propagation of the THz pulse. For angles in between 0 and 90$^\circ$, the polarization state is generically elliptically polarized, where the phase difference between the orthogonal linearly polarized components is not $\pm\pi/2$, and their amplitudes are not equal to each other. At exactly 45$^\circ$, i.e. at the middle point between the orientations that produce circular polarization, the polarization state of the THz beam propagating through the array remains linear.  This means that we can keep the array in the THz spectrometer and just adjust its orientation to generate linear or circular polarization states depending on the need of the experiment, as shown in figure \ref{setup}.

Combining the polarization modulation technique with the Fresnel rhomb array allows the technique to show its true power. Importantly, we can measure directly the elements of the $T$-matrix in the circular polarization basis (table 1) by using a pair of Fresnel rhomb arrays in the configuration
\begin{eqnarray}
E_{out}=P_y\cdot QWP_{-\pi/4}\cdot T\cdot QWP_{-\pi/4}\cdot P_{\Omega t}\cdot E_y
\label{eq:Eout_full}.
\end{eqnarray} 
This results in 
\begin{eqnarray}
X=-\tfrac{1}{4}\left(a-d-i(b+c)\right)=-\tfrac{1}{2}t_{+-},
\end{eqnarray} and
\begin{eqnarray}
Y=-\tfrac{i}{4}\left(a+d+i(b-c)\right)=-\tfrac{i}{2}t_{++}
\end{eqnarray}
which are the elements in the top row of the $T$-matrix in circular basis, as shown in table 1. Correspondingly, the configuration 
\begin{eqnarray}
E_{out}=P_y\cdot QWP_{+\pi/4}\cdot T\cdot QWP_{-\pi/4}\cdot P_{\Omega t}\cdot E_y
\end{eqnarray}
results in
\begin{eqnarray}
X=\tfrac{1}{4}\left(a+d-i(b-c)\right)=\tfrac{1}{2}t_{--},
\end{eqnarray} and
\begin{eqnarray}
Y=\tfrac{i}{4}\left(a-d+i(b+c)\right)=\tfrac{i}{2}t_{-+}
\end{eqnarray} 
which are the elements in the bottom row of the circular basis $T$-matrix, also shown in table 1. Therefore, just as in the linear polarization basis, we need only two measurements to completely characterize a sample, but now in the circular basis.

\section{Results}

\begin{figure}[t!b!h!]
\includegraphics[width=.9\columnwidth]{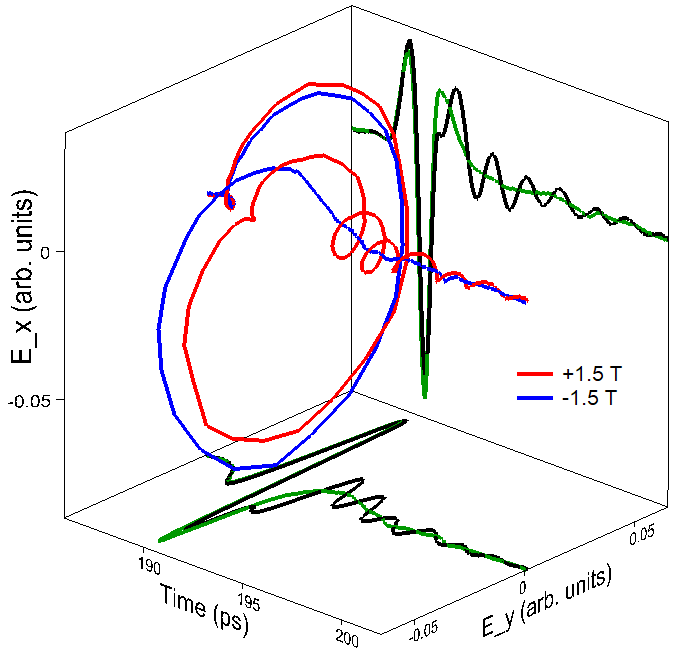}
\centering
\caption{Time dependence of transmitted electric field through GaAs--AlGaAs quantum well at $\pm$ 1.5 T with incident right circularly polarized light. Time oscillations are clearly observed only for positive magnetic field, as expected.}
\label{fig:twodeg}
\end{figure}

\begin{figure*}[t!b!h!]
\includegraphics[width=.9\textwidth]{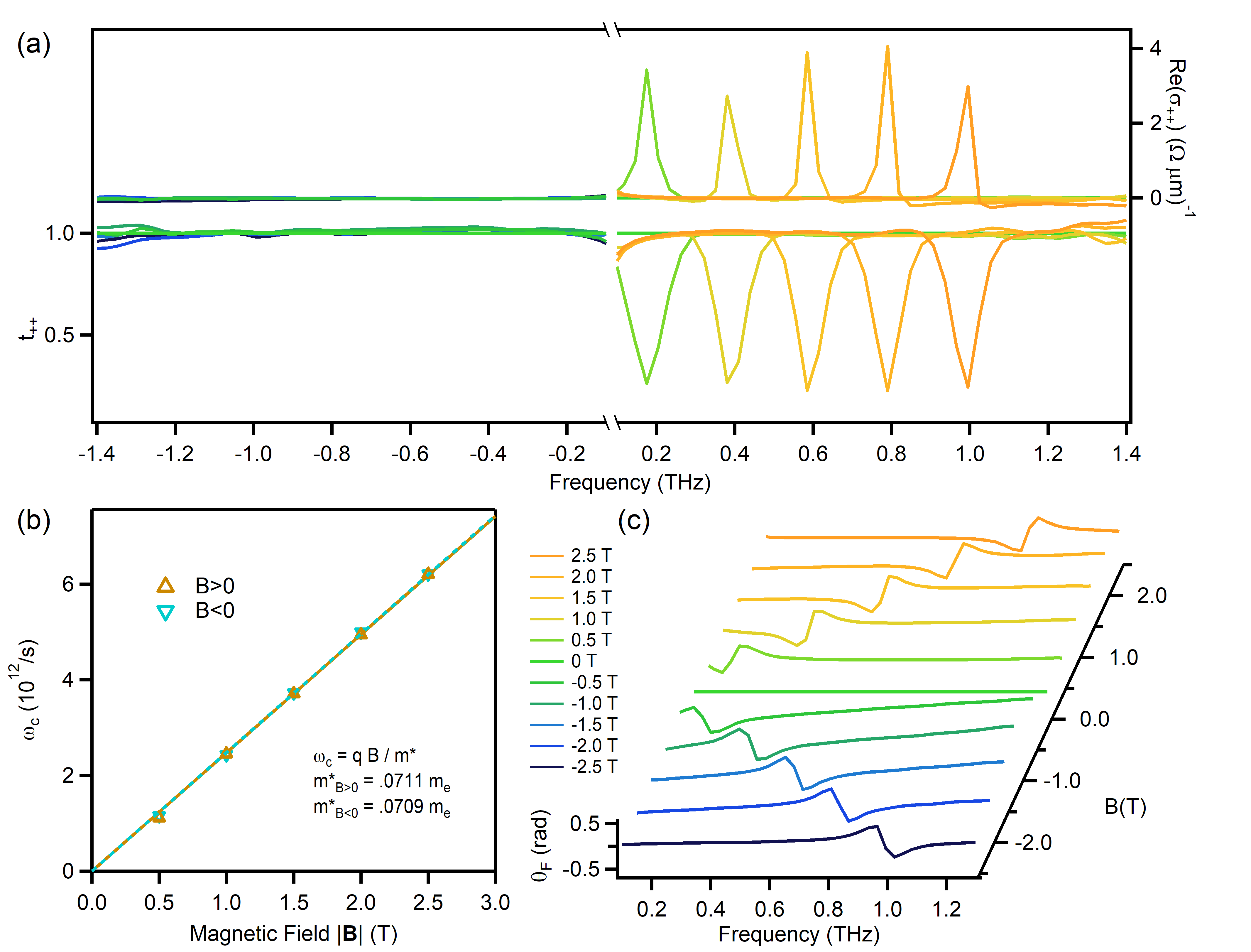}
\centering
\caption{(a) $t_{++}$ and $\sigma_{++}$ of the AlGaAs-GaAs quantum well with applied magnetic field values as a function of frequency. Negative frequency is shown here to represent negative magnetic field values. Resonant absorption occurs only at the cyclotron resonance frequency. We have omitted the frequencies below 100 GHz due to low signal-to-noise ratio. (b) Cyclotron frequency versus magnetic field, clearly showing a linear behavior; the effective mass is extracted to be 0.071 m$_e$. (c) Faraday rotation from the cyclotron resonance for fields from -2.5 to 2.5 T.}
\label{tcef}
\end{figure*}

We now show the versatility of the combination of the Fresnel rhomb array and the polarization modulation technique by studying the cyclotron resonance in a sample of AlGaAs--GaAs quantum well. It is well known that the cyclotron resonance response of an electronic system is a consequence of the Hall effect that charge carriers undergo in the presence of an external magnetic field \cite{PalikFurdyna}. In the semi-classical approximation, the optical response due to free carriers in the presence of a magnetic field for an isotropic material (and for a four-fold symmetric one as well), is given by the circular polarization conductivities 
\begin{eqnarray}
\label{conductivity}
\sigma_\pm=\frac{\omega_p^2}{4\pi(\gamma-i(\omega\mp\omega_c))}.
\end{eqnarray}
In this expression, $\omega_p^2=\tfrac{nq^2}{m^*}$ is the plasma frequency and is a measure of how many charge carriers participate in the conductivity, $\gamma$ is the transport scattering rate (also known as the momentum relaxation rate), and $\omega_c=\tfrac{qB}{m^*}$ is the cyclotron angular frequency. Here $\sigma_+$ and $\sigma_-$ are the response to right and left circularly polarized radiation, respectively. They are also called cyclotron resonance active (CRA) and cyclotron resonance inactive (CRI) conductivities respectively because they correspond to a positive and negative cyclotron frequency. One can obtain the same effect by changing the direction of the magnetic field, as this would change the sign of the cyclotron frequency. Therefore, for a given circular polarization of the incoming THz pulse into the sample, only one direction of the magnetic field would show the cyclotron resonance (CRA), while the opposite direction of the field would allow the pulse to transmit without change and it would not show any resonance (CRI).
%
%
This behavior is displayed in figures \ref{fig:twodeg} and \ref{tcef}. Figure \ref{fig:twodeg} shows the effect of the cyclotron resonance activity in an impinging right circularly polarized THz pulse. For one direction of the field (-1.5 T) the resonance is present in the form of periodic oscillations of the amplitude of the response. For a field in the opposite direction (+1.5 T), the pulse traverses the sample unaffected and without any temporal oscillations, indicating the absence of cyclotron resonance absorption. Panel (a) of figure \ref{tcef} shows the CRA and CRI transmission magnitude for magnetic field values from -2.5 T to 2.5 T referenced to the response of the sample at zero magnetic field. This corresponds to the upper left matrix element in the circular basis $t_{++}$. The position of the dip in $t_{++}$ corresponds to the cyclotron frequency. Panel (a) also shows the real part of the thin film conductivity extracted from the transmission, $T$, according to the formula 
\begin{eqnarray}
T(\omega) = \frac{1+n_{GaAs}}{1+n_{GaAs}+Z_0 \sigma(\omega) d},
\end{eqnarray}
where $n_{GaAs}$ is the index of refraction of the GaAs substrate, $Z_0$ is the impedance of free space, and $d$ is the thickness of the quantum well layers on the substrate (estimated here as 10 nm). The conductivity shows spikes at the cyclotron frequencies because the real part of the conductivity is the lossy part while the imaginary conductivity is lossless. By fitting the conductivity with the formula \ref{conductivity} for the circular polarization conductivity, we can extract the cyclotron frequency ($\omega_c$) as a function of magnetic field. We have plotted $\omega_c$ for several magnetic field values in panel (b). Using a linear fit for the cyclotron frequency, we find from the slope the cyclotron mass of 0.071 m$_e$. Finally, we obtain the Faraday rotation angle from the formula 
\begin{eqnarray}
\theta_F=\tfrac{1}{2}\arg{\tfrac{t_{--}}{t_{++}}}
\end{eqnarray}
and show it in panel (d). The cyclotron frequency is also evident here as the point of maximum slope in the Faraday angle.

\section{Summary}
We developed a method in TDTS whereby the full frequency-dependent transmission matrix can be obtained directly in the circular polarization basis. This method required two broadband QWPs; one wave plate encodes the THz from the linear basis to the circular polarization basis and the other decodes it back to the linear polarization basis where we can detect it with conventional means. Utilizing TOPAS\textsuperscript{\textregistered}, a polymer with a flat index of refraction and near zero absorption, we designed, built, and tested two Fresnel rhomb arrays. Each array performed well and provided a phase shift to the THz of $\sim\pm\pi/2$ for frequencies below $\sim$ 1.5 THz. We tested our technique by obtaining the circular polarization basis transmission matrix in a 2DEG in a AlGaAs-GaAs quantum well, which responds strongly to only one handedness of input light. From the 2DEG we extracted $\sigma(\omega)$, $\theta_F(\omega)$, and m$^*$ = .071 m$_e$. This technique represents a significant advance in the ability to perform intuitive experiments on materials that break time-reversal symmetry.

\begin{acknowledgments}
We acknowledge technical assistance from E. Johnston-Halperin and R. Kawakami at OSU. This work was partially supported by OSU's Institute for Materials Research under grants IMR-GF0155 and EMR-G00019, and by the Center for Emergent Materials, an NSF MRSEC, under grant DMR-1420451.
\end{acknowledgments}

\bibliography{PolModQWP}

\end{document}